\documentclass[10pt]{iopart}
\usepackage{microtype}
\usepackage{iopams}
\usepackage{graphicx}
\usepackage{cite}
\usepackage{color}

\usepackage{epsfig}

\expandafter\let\csname equation*\endcsname\relax 
\expandafter\let\csname endequation*\endcsname\relax 

\usepackage{amsmath}
\usepackage{bm}
\usepackage{color}

\newcommand{\col}[1]{\textcolor[rgb]{0,0,0}{#1}}

\newcommand{\be}{\begin{equation}}
\newcommand{\ee}{\end{equation}}
\newcommand{\bea}{\begin{eqnarray}}
\newcommand{\eea}{\end{eqnarray}}

\def\<{\langle}
\def\>{\rangle}

\def\lam{\lambda}

\usepackage{amsmath,amsthm,amstext,amscd,amssymb,euscript,mathrsfs}
\usepackage{calrsfs}
\usepackage{epsfig}
\usepackage{color}
\usepackage{url}

\newcommand{\R}{\mathbb R}

\newcommand{\E}{\mathbb E}

\renewcommand{\phi}{\varphi}

\def\1{{\mathchoice {\rm 1\mskip-4mu l} {\rm 1\mskip-4mu l}
{\rm 1\mskip-4.5mu l} {\rm 1\mskip-5mu l}}}

\newtheorem*{theorem*}{{\small T}{\scriptsize HEOREM}}
\newtheorem{corollary}{{\bf{\small C}{\scriptsize OROLLARY}}}[section]
\newtheorem*{proposition*}{{\bf{\small P}{\scriptsize ROPOSITION}}}
\newtheorem{lemma}{{\bf{\small L}{\scriptsize EMMA}}}[section]

\newtheorem{definition}{{\bf{\small D}{\scriptsize EFINITION}}}[section]

\renewenvironment{proof}[1]
{\noindent{{\bf{\small{ P}{\scriptsize ROOF}}}.}\hspace{0.1cm} #1} 

\newcommand{\beq}{\begin{eqnarray}}
\newcommand{\eeq}{\end{eqnarray}}

\newcommand{\ba}{\begin{align*}}
\newcommand{\ea}{\end{align*}}

\newcommand{\bl}{\begin{lemma}}
\newcommand{\el}{\end{lemma}}

\newcommand{\bt}{\begin{theorem}}
\newcommand{\et}{\end{theorem}}

\newcommand{\bd}{\begin{definition}}
\newcommand{\ed}{\end{definition}}

\newcommand{\bc}{\begin{corollary}}
\newcommand{\ec}{\end{corollary}}

\newcommand{\bpr}{\begin{proof}}
\newcommand{\epr}{\end{proof}}

\newcommand{\ben}{\begin{enumerate}}
\newcommand{\een}{\end{enumerate}}

\newcommand{\caC}{{\mathscr C}}

\begin{document}

\title{Spatial fluctuation theorem}

\author{Carlos P\'erez-Espigares,$^1$ Frank Redig,$^2$ and Cristian Giardin\`a$^1$}
\address{
$^1$\ University of Modena and Reggio Emilia, via G. Campi 213/b, 41125 Modena, Italy\\
$^2$\ University of Delft, Mekelweg 4 2628 CD Delft , The Netherlands
}
\eads{
\mailto{carlos.perezespigares@unimore.it},
\mailto{F.H.J.Redig@tudelft.nl},
\mailto{cristian.giardina@unimore.it}
}



\date{\today}

\begin{abstract}
{For non-equilibrium systems of interacting particles and for interacting diffusions in $d$ dimensions, 
a novel fluctuation relation is derived.
The theorem establishes a quantitative relation between the probabilities of observing two current values 
in  different spatial directions.
The result is a consequence of {\em spatial symmetries} of the microscopic dynamics, generalizing in this 
way the Gallavotti-Cohen fluctuation theorem related to the time-reversal symmetry. This new perspective opens up
the possibility of direct experimental measurements  of fluctuation relations of vectorial observables.
}
\end{abstract}

\pacs{02.50.-r, 05.70.Ln, 05.40.-a, 66.10.C--}

\vspace{2pc}
\noindent{\it Keywords}: Fluctuation theorem, Large deviations, Nonequilibrium and irreversible thermodynamics


\noindent
\section{Introduction}

The understanding of non-equilibrium systems has been a main goal of statistical physics over the last  decades,
where a remarkable progress has been reached. 
Key to this advance has been the study of the statistics of several observables ranging 
from the entropy production or the  work, to the density or the current (of particle, charge, energy, momentum...).
As the probability distribution of these observables follow a large deviation principle, the study of their corresponding large 
deviation functions (LDFs) has become a subject of primary interest, because LDFs can be considered as the {\em free energy} analog
in non-equilibrium systems \cite{derrida2007,Touchette}. For instance, the Macroscopic Fluctuation Theory \cite{bertini2014},  provides a  
\col{variational} formula from which
one can compute the LDF of joint space-time densities and currents for driven diffusive systems, just by knowing two transport coefficients.

Particularly important in the understanding of non-equilibrium large deviations 
are the so called {\em fluctuations relations} 
\cite{EvansFT,GCFT,JarzEq,KurchanFT,LSFT,MaesFT, MaesRedig,Crooks1999,seifert2005,harris2007,chetrite2008,garcia2012}, 
valid arbitrarily far from equilibrium. Those relations  
basically exploit the {\em time reversal symmetry} of the
microscopic dynamics, 
that gives rise to a relation between a positive fluctuation 
of certain observable and the corresponding negative fluctuation.
However, for systems endowed with a spatial structure one may wonder if  {\em spatial symmetry} yields relations between the probabilities of vectorial observables in different directions. This was  pointed out in \cite{IFR}, where 
an Isometric Fluctuation Relation (IFR) was obtained within the Macroscopic Fluctuation Theory framework under some assumptions and verified using numerical simulations. 
The IFR relation allows us to relate in a very simple manner any pair of isometric current fluctuations
and it has been recently extended to the context of anisotropic systems \cite{AFR}.
In the IFR derivation the two dimensional system at hand was driven out of equilibrium by boundary reservoirs at different values in the presence of a field. In the present paper we consider systems driven far from equilibrium in the presence of a bulk field, i.e bulk-driven systems, and derive fluctuation theorems as a consequence of general spatial transformations, thereby extending the IFR case which was restricted to rotations.

Fluctuations relations are very relevant because from them one can derive in the linear response regime 
the Onsager reciprocity relations and the Green-Kubo formulas and, even more important, other reciprocity relations beyond these two can be obtained by considering higher order response terms \cite{AndGasp}.
In a similar way the IFR relation implies a set of new hierarchies between current cumulants
and non-linear response coefficients \cite{IFR}.

The focus of this paper is on deriving {\em spatial fluctuation theorems} starting
from the underlying stochastic microscopic dynamics. Whereas a clear understanding 
has been reached for the microscopic origin of the standard fluctuation
theorem \cite{GCFT,KurchanFT,LSFT,MaesFT}, it is an open problem to relate spatial 
fluctuation theorems to invariance properties of the microscopic dynamics.
To achieve the result we will employ the Gibbs-formalism in space-time as introduced in 
\cite{MaesFT}. 
The fluctuation symmetries which we derive here are a consequence of this space-time Gibbsian structure. They apply both in equilibrium context, when the space-time Hamiltonian is time-reversal invariant 
(see \cite{lacoste} for the analysis {\em at equilibrium})
as well as in systems driven away from equilibrium, for which time reversal symmetry is broken.

\medskip

\section{Discrete systems}
We first consider particle systems embedded in $\mathbb{Z}^d$  following 
a Markov dynamics. For the sake of simplicity we restrict to  
discrete time, however similar results hold true for continuous
time dynamics.
The Markov process is denoted by $\{X(n) = (X_{k}(n))_{k\in \mathbb{N}}  :\; n \in \mathbb{N}\}$,
where $X_{k}(n)$ denotes the random position of the $k^{th}$ particle at time $n$.
System configurations are denoted by the vector $x=(x_1,x_2,\ldots)$ with the
first particle located at site $x_1\in \Lambda_L = \{1,\ldots, L\}^d \subseteq \mathbb{Z}^d$ 
and similarly for all the other particles.
\col{During the evolution particles 
jump to their nearest neighbor sites with anisotropic probabilities}.
The anisotropy is tuned by the  $d$-dimensional
vector $a=(a_1,\ldots,a_d)$ and in each direction we assume 
a weak asymmetry which is produced by a constant  external field
$E= (E_1,\ldots,E_d)$. The dynamics is defined by the
probability transition matrix with elements 
\be
\label{transition}
p(x,y)  
= \left\{
\begin{array}{ll}
\frac{b(x,y)}{C} a_s e^{\pm \frac{E_s}{L}} &  \textrm{if } y = x^{k,\pm e_s}\\
0 & \textrm{otherwise,}  
\end{array} \right.
\ee
for $k\in\mathbb N$ and $s =1,\ldots, d$.
Here  $b(x,y)$ is a generic transition kernel 
from configuration $x$ to configuration $y$, $C$ is the normalization constant,
$e_s$ is the unit vector in the $s$-direction and
$x^{k,\pm e_s}$ is the  configuration that is obtained from
the configuration $x$ by moving the $k^{th}$ particle  to
$x_k \pm e_s$. 

We consider an initial particle number (\col{a conserved quantity of the dynamics})
proportional to the volume, i.e. $N= \rho L^d$, corresponding to a constant density $0 < \rho <\infty$.
{We will be interested in the large deviations of the current vector 
$Q_{M,L}  =  \frac{1}{L}(N^{+,1}-N^{-,1},\dots,N^{+,d}-N^{-,d})$, where $N^{\pm,s}$ are the number of particle 
jumps in the $\pm s$ direction up to time $M$.}
This current 
is  an
additive functional of the Markov process $(X(n))_{n\in\mathbb{N}}$ and 
it satisfies a large deviation principle that can be informally stated as
$
\mathbb{P}(Q_{M,L} \sim M q) \approx e^{-M I_L(q)}
$
as $M\to\infty$.
By the Gartner-Ellis theorem \cite{Den-Hollander} the large deviation function $I_L$ can be 
obtained as
$
I_L(q) = \sup_{\lambda} ( \lambda \cdot q - \mu_{L}(\lambda))
$	
where we define the 
scaled cumulant generating function 
of the current as 
\be
\mu_{L}(\lambda) := \lim_{M\to\infty} \frac{1}{M} \ln Z_{M,L}(\lam),
~~\text{with}~~Z_{M,L}(\lam)=\mathbb{E}(e^{\lambda \cdot Q_{M,L}}),
\nonumber
\ee
where $\mathbb{E} (\cdot)$ 
denotes the average in path space.
The application of Perron-Frobenius theorem allow us to express
$\mu_L(\lambda)$ as {the logarithm of the largest
eigenvalue of the tilted matrix}  with elements 
$p_\lambda(x,y)=b(x,y) a_s e^{\pm \left(  \frac{E_s}{L}+\frac{\lambda_s}{L} \right)}$
if $y = x^{k, \pm e_s}$ and $p_\lambda(x,y)=0$ otherwise.
To study the system in the thermodynamic limit one needs to rescale by defining 
\be
\label{defi}
\mu(\lambda) := \lim_{L\to\infty} \frac{\mu_{L}(\lambda)}{L^{d-2}} =  \lim_{L\to\infty} \lim_{M\to\infty} \frac{\ln Z_{M,L}(\lam)}{L^{d-2}M}. 
\ee

Our first result provides the derivation of an {\em anisotropic fluctuation relation} in $d$ dimensions 
from the assumption of a system invariance property.
Consider $N$ interacting particles 
whose dynamics is defined by 
the transition matrix \eqref{transition}.
We denote by $P$  the probability measure on the path space
$\Omega$ (i.e. the space of all trajectories) in the presence of an external  field $E=(E_1,\dots,E_d)$,
and $P_0$  the corresponding measure with $E=0$.
\begin{theorem*}[Spatial fluctuation theorem]
Let $U:\mathbb{Z}^d \to \mathbb{Z}^d$ be a  transformation on the physical space.
For a trajectory in path space $\vec{x} = (x(0),x(1),\ldots x(M))$, 
consider the bijective mapping ${\cal R}: \Omega \to \Omega$ induced by 
$U$: 
$
({\cal R}{x}(n))_{k} = U{x}_{k}(n)\;.
$
Assume that the current satisfies
\be
\label{assume2}
Q_{M,L}({\cal R}\vec{x}) = U \, Q_{M,L}(\vec{x})
\ee
and assume that $P_0$ has the invariance property
\be
\label{invariance-prop}
{P}_{0}(\vec{x})={P}_{0}({\cal R} \vec{x})  \qquad \forall \; \vec{x} \in \Omega\;.
\ee
Then the following fluctuation relations hold: $\forall \lambda$
\be
\label{afr-finite-vol}
Z_{M,L}(\lam) = Z_{M,L}\Big((U^{-1})^t (\lam + E) - E\Big),
\ee
\be
\label{I2-finite-vol}
I_{L}(q) -I_L(Uq)= (U^t E-E) \cdot q
\ee
where $(U^{-1})^t$ denotes the transposed of the inverse.
\end{theorem*}
\col{Notice that statement \eqref{afr-finite-vol} is obtained at finite time and finite volume,
whereas eq. \eqref{I2-finite-vol} holds for any finite volume in the limit of large times.
Furthermore, when the limit in \eqref{defi} is finite, one has $\mathbb{P}(Q_{M,L} \sim M L^{d-2}q) \approx e^{-M L^{d-2}I(q)}$
with $I(q)=\lim_{L\to \infty} I_L(q)/L^{d-2}$ satisfying the analogous of eq. \eqref{I2-finite-vol}.}
We remark as well  that to satisfy relation \eqref{invariance-prop} the map $U$ will also 
depend on the volume $L$ and the anisotropy $a$. However, to alleviate
notation we do not write this dependence explicitly.
In the proof 
we also shorthand $Z_{M,L}(\lam)$ as $Z(\lam)$ and $Q_{M,L}(\vec x)$ as $Q(\vec x)$.

\smallskip 

\begin{proof}
We start by observing that, from the previous definition of the current 
it holds that ${P}(\vec{x})=\exp{[E \cdot Q(\vec{x})]}{P}_0(\vec{x})/{\cal N}$, with 
${\cal N}=\sum_{\vec{x}\in\Omega}\exp{[E \cdot Q(\vec{x})]}{P}_0(\vec{x})$. 
Applying this relation 
we have
\bea
Z(\lambda)
= \sum_{\vec{x}\in\Omega}{P}(\vec{x}) e^{\lambda \cdot Q (\vec{x})} = \sum_{\vec{x}\in\Omega}\frac{{P_0}(\vec{x})}{{\cal N}} e^{(E+\lambda) \cdot Q (\vec{x})}.
\nonumber
\eea
Thus, by using the invariance property \eqref{invariance-prop} we get
\bea
Z(\lambda)
=\sum_{\vec{x}\in\Omega}\frac{{P_0}({\cal R} \vec{x})}{{\cal N}} e^{(E+\lambda) \cdot Q (\vec{x})}=
\sum_{\vec{x}\in\Omega}{P}({\cal R} \vec{x}) e^{-E\cdot Q ({\cal R}\vec{x})+
(E+\lam) \cdot Q(\vec{x})}
\nonumber
\eea
Applying the change of variables $\vec{y}= {\cal R}\vec{x}$ and since ${\cal R}$ is a bijective map 
we find
\bea
Z(\lambda) = \sum_{\vec{y\in \Omega}}{P}(\vec{y}) e^{-E\cdot Q (\vec{y})+
(E +\lam) \cdot Q({\cal R}^{-1}\vec{y})} .\qed
\nonumber
\eea
Hence, using  the assumption \eqref{assume2}, it is easy to check that \eqref{afr-finite-vol} follows.
By taking the limit $M\to \infty$ the same relation holds for $\mu_L$ and therefore \eqref{I2-finite-vol} follows by Legendre transform.
\end{proof}
\section{Comments and examples}
If  the transformation $U$ is chosen as spatial inversion, i.e. $U i = - i$ for $i\in\mathbb{Z}^d$, then
one recovers the standard Gallavotti-Cohen 
fluctuation relation, i.e.
$
\mu_L(\lam) = \mu_L(-\lam-2E).
$
Notice that usually this relation is associated to time reversal invariance 
of the  measure of the symmetric system.
It  was  remarked in \cite{LSFT} that any transformation on path space
such that ${\cal R} \circ {\cal R} = 1$ would lead to the Gallavotti-Cohen fluctuation
relation. The transformation on path space induced by 
spatial inversion in physical space has indeed such property.

\begin{figure}[h]
\centering
\includegraphics[scale=0.7]{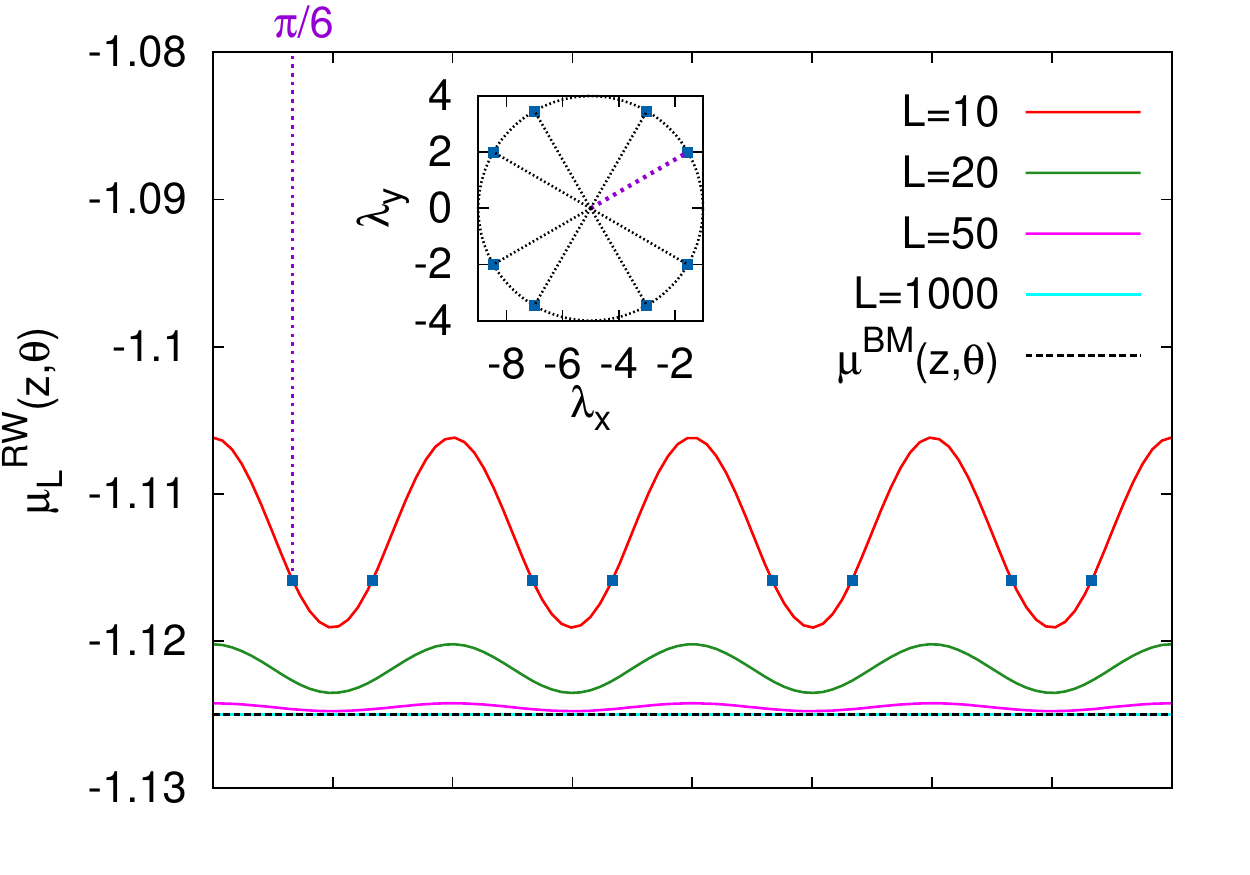}\vspace*{-6mm}\\
\includegraphics[scale=0.7]{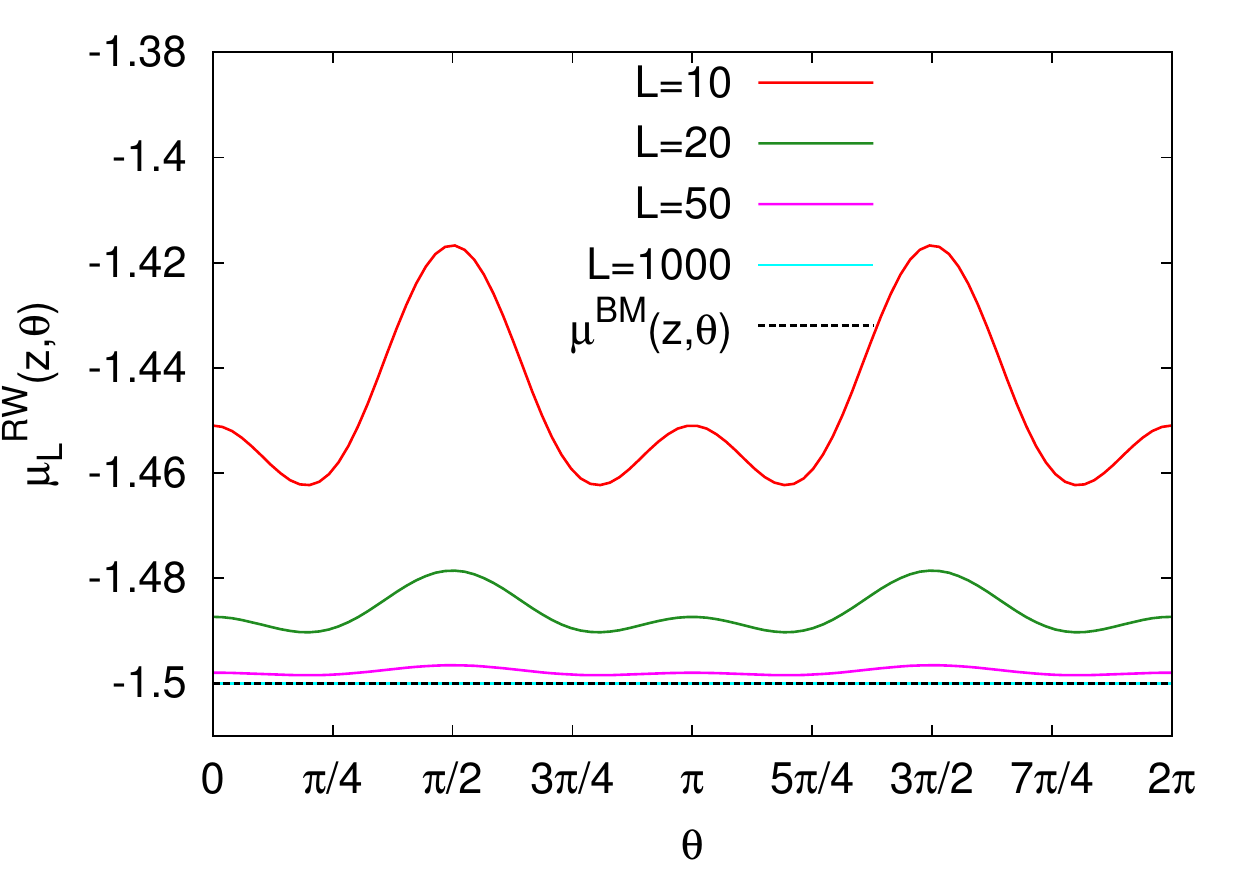}
\caption{(Color online) Plot of $\frac{\mu_L^{RW}(z,\theta)}{L^{d-2}}$ in $d=2$
as a function of $\theta = \arctan ( \frac{1}{\sqrt{\alpha}} \frac{\lambda_2 + E_2}{\lambda_1+E_1})$.
In both figures $\rho=0.5$, $E = (5,0)$, $z = 4$ and
the anisotropy ratio is $\alpha=1$ in the top and  $\alpha=2$ in the bottom.
The inset of the top panel shows the 8 discrete symmetries of the
system in $\lambda$-space for $L=10$, starting from an initial angle e.g. $\theta=\pi/6$, corresponding to the transformations $U$ 
given by the diagonal 
and the anti diagonal matrices with elements $\pm 1$.
As $L$ increases the discrete symmetry  scales to a continuous symmetry
associated to the transformation 
$
U=\left( {\begin{array}{cc}
   \cos\theta & -\sqrt{\alpha} \sin\theta \\
    \frac{1}{\sqrt{\alpha}} \sin\theta & \cos\theta \\
  \end{array} } \right)
$
and \eqref{afr-infinite-vol} is then satisfied.
}
\label{anisofig}
\end{figure}

More generally, the theorem above allows us to deduce generalised fluctuation relations 
as a consequence of {\em spatial symmetries}, i.e. whenever a transformation
$U$ on the physical space satisfies \eqref{assume2}  and \eqref{invariance-prop}
then \eqref{afr-finite-vol} and \eqref{I2-finite-vol} follow. 
To further illustrate this point we shall discuss examples of systems of non-interacting particles.
In this case 
the dynamics can be studied in terms of a single particle
and the scaled cumulant generating
function 
$\mu_L$ can be explicitly solved, so that one can check
by inspection which spatial symmetries hold.
For instance, for a system of independent random walkers (RW) where
each particle at site $i$ jumps to site $i\pm e_s$
with probability $a_s e^{\pm\frac{E_s}{L}}$ with periodic
boundary conditions, an elementary
application of Perron-Frobenius theorem gives
\be
\mu^{\textrm{RW}}_L(\lam)= \rho L^d \ln\left[ \sum_{s=1}^d \left(a_s e^{(E_s+\lam_s)/L} + a_s e^{-(E_s+\lam_s)/L}\right)   \right].
\nonumber
\ee
For $d=2$ and by doing 
a change of variables to polar coordinates ($z$, $\theta$) 
such that $\lam_1=z \cos\theta -E_1$
and $\lam_2=z \sin \theta \sqrt{\alpha} -E_2$, with $\alpha=a_1/a_2$ being the anisotropy 
ratio, the above expression reads
\bea
\mu^{\textrm{RW}}_L(z,\theta)&=& \rho L^2 \ln\left[ a_1 \left( e^{\frac{z\cos\theta}{L}}+e^{\frac{-z\cos\theta}{L}} \right)\right.\nonumber\\ 
&&\left. +a_2 \left( e^{\frac{z\sqrt{\alpha}\sin\theta }{L}}+e^{\frac{-z\sqrt{\alpha}\sin\theta }{L}} \right)   \right].
\label{thetaIRW2}
\eea
\col{In the isotropic case ($\alpha=1$) one recognizes by inspection 
the {\em discrete symmetries} of the system leading to a fluctuation theorem.}
Namely, $\mu^{\textrm{RW}}_L(z,\theta)=\mu^{\textrm{RW}}_L(z,\theta')$,  
for $\theta'=m \pi/2\pm \theta$, $\forall m\in \mathbb Z$ (see Fig. \ref{anisofig}). 
It is then natural to expect that for diffusions
one is lead to consider {\em continuous spatial symmetries}.
This can be seen by considering the diffusive scaling limit of
the previous example.
If we define  
\col{$\{X^{(L)}(n) = (X^{(L)}_{k}(n))_{k\in \mathbb{N}}  :\; n \in \mathbb{N}\}$}
the positions \col{at time $n$} of the $N$ independent random walkers (labeled by $k$) in a volume of linear 
size $L$,
then
the process 
$\{R_t = (R_{k,t})_{k\in \mathbb{N}}  :\; t \in \mathbb{R}^+\}$
defined by 
$
R_t := \lim_{L\to \infty} \frac{X^{(L)}(\lfloor L^2 t \rfloor)}{L}
$
will be a family of $N$ independent anisotropic Brownian motions  
satisfying 
the stochastic differential equation
$
dR_{k,t} =  A E dt + \sqrt{A} dB_{k,t} 
$
with $A$ the 
$d\times d$
diagonal diffusion matrix with elements $A_{s,s}=2 a_s$ and  
$B_{k,t} \in \R^d$
denoting
independent standard Brownian motions.
An immediate computation gives
\be
\label{mubm}
\lim_{L\to\infty} \frac{\mu^{RW}_{L}(\lambda)}{L^{d-2}} = \mu^{BM}(\lambda) =
\rho\sum_{s=1}^d a_s \lambda_s(\lambda_s + 2E_s)
\ee
where 
$
\mu^{\textrm{BM}}(\lam) := \lim_{T\to \infty} \frac{1}{T L^d} \ln \Big[\mathbb{E}(e^{\lambda \cdot Q_T})\Big]
$,
being $Q_T=\sum_{k=1}^N R_{k,T}$ the current up to time $T$.
From the explicit  expression \eqref{mubm} we see that a spatial fluctuation relation 
holds, i.e.
\be
\label{afr-infinite-vol}
\mu^{BM}(\lam) = \mu^{BM}\Big((U^{-1})^t (\lam + E) - E\Big) 
\ee
with $U$ such that
$
\label{cond-U}
U A U^t = A
$.
For $d=2$, $U$ takes the form given at the end of the caption of Fig. \ref{anisofig}.
As we shall see below, such a relation can be traced back to the invariance of the path measure
of the anisotropic Brownian motion under a spatial transformation $U$
such that $U A U^t = A$
,in agreement with the findings of \cite{IFR,AFR} in the context of the 
Macroscopic Fluctuation Theory \cite{bertini2014}. 
Notice indeed that the MFT predicts for a two dimensional anisotropic periodic system in the presence of a filed $E$ that 
the probability of having a time-integrated current fluctuation $q_0=\frac{1}{T}\int_0^T dt \int dr q(r,t)$ up to a long time $T$ is 
$P(q_0)\sim e^{-T L^d I(q_0)}$ with the LDF \cite{AFR} (see also \cite{PRECarlos} for the one dimensional case)
\be
\label{LDFMFT}
I(q_0)=\lim_{T\to \infty}\inf_{(\rho,q)\in \Omega}\frac{1}{T}{\cal I}(\rho,q),
\ee
where
\be
\nonumber
{\cal I}(\rho,q)= \frac{1}{2}\int_0^T dt\int dr (q+D(\rho)\nabla \rho -E \sigma(\rho))^T\sigma(\rho)^{-1}
(q+D(\rho)\nabla \rho -E \sigma(\rho)).
\ee
Here the diffusivity and the mobility are given by $D(\rho)$ and $\sigma(\rho)$ respectively.
In \eqref{LDFMFT}, $\Omega$ is the set of paths $(\rho,q)$ satisfying the continuity equation and whose 
time-integrated current is $q_0$,
\be
\nonumber
\Omega=\left\{ (\rho,q):\frac{1}{T}\int_0^Tdt\int dr q(r,t)=q_0, \partial_t \rho=-\nabla\cdot q \right\}
\ee
The cumulant generating function is then given by the Legendre transform of \eqref{LDFMFT}
\be
\mu(\lam)=\sup_{q_0}[\lam q_0-I(q_0)].
\ee
In particular, for $N$ anisotropic independent random walkers in a volume $L^d$ such that $\rho=N/L^d$, 
$D(\rho)$ and $\sigma(\rho)$ are given by the diagonal matrices with elements $a_s$ and $2\rho a_s$ respectively. 
In this case, it is easy to check that by solving the variational problem given by \eqref{LDFMFT} we get \eqref{mubm} as expected.

\medskip

\section{Diffusions}
To state a (spatial) fluctuation relation 
for systems following a generic diffusion process
we consider an abstract path space $\Omega$, 
and a bijective measurable transformation
${\cal R}:\Omega\to\Omega$ with inverse ${\cal R}^{-1}$.
Elements in the path space are denoted by $\omega$.
For diffusions we will consider
$\Omega= \caC([0,T],\R^d)$, i.e. the set
of all continuous paths up to time $T$ taking value in $\R^d$,
and
$
({\cal R}(\omega))_t= U \omega_t
$,
with $U:\R^d\to\R^d$ an invertible spatial transformation.
\begin{proposition*}
Consider a probability measure on path space $\Omega$ of the form
$
P(d\omega)= e^{H(\omega)} P_0(d\omega)
$
where $P_0$ is ${\cal R}$-invariant.
For all $\phi:\Omega \to \mathbb{R}$ we have the identity
\be\label{fluctgen}
\E(e^{\phi\circ {\cal R}})
=  \E(e^{\phi} e^{H\circ {\cal R}^{-1}- H})
\ee
\end{proposition*}
\begin{proof}
\begin{eqnarray*}
\int e^{\phi\circ {\cal R}} dP &=& \int e^{\phi\circ {\cal R}}e^H dP_0 =
\int e^{\phi}e^{H\circ {\cal R}^{-1}} dP_0 \\
&=&\int e^{\phi}e^{H\circ {\cal R}^{-1}- H} dP.\qed
\end{eqnarray*}
\end{proof}
\noindent In the particular case that $\phi= -\gamma(H\circ {\cal R}^{-1}-H)$ \col{with $\gamma\in\mathbb{R}$}, the relation \eqref{fluctgen} gives
$
\E(e^{\gamma (H\circ {\cal R}-H)})= \E (e^{(1-\gamma)(H\circ {\cal R}^{-1}-H)}).
$
Even more, specifying to a transformation such that ${\cal R}={\cal R}^{-1}$ this
identity is exactly a symmetry of the form of the standard fluctuation theorem for the quantity
$H\circ {\cal R}-H$.
\\
Denoting by $P\circ {\cal R}$ the image measure of $P$ under ${\cal R}$, 
i.e.,
$
\int f d(P\circ {\cal R}) = \int (f\circ {\cal R} ) dP,
$
it can be readily verified that $dP=e^{H} dP_0$ implies that $d(P\circ {\cal R})= e^{H\circ {\cal R}^{-1}}dP_0$.
Therefore, as we shall use below, we can write
\be\label{alt}
e^{H\circ {\cal R}^{-1}- H}=  \frac{d(P\circ {\cal R})}{dP}= \frac{d(P\circ {\cal R})}{dP_0}\Big/ \frac{dP}{dP_0}.
\ee

The abstract setting of the proposition above can be used to 
derive the spatial fluctuation theorem for finite time $T$ in the context of 
interacting diffusions.
We shall illustrate this by considering the overdamped Langevin dynamics 
$\{X_t = (X_{k,t})_{k\in \mathbb{N}}  :\; t \in \mathbb{R}^+\}$ 
describing $N$ particles (labeled by $k$) subject to a drift vector 
which can arise from an applied force to each particle $F$ and/or a conservative potential $V$, with a 
positive definite constant diffusion matrix $A$. Then the stochastic differential equation for 
the $k^{th}$ particle reads (Ito convention)
\be
\label{origpr}
dX_{k,t}= F(X_{k,t}) dt + \nabla_k V(X_t) dt + \sqrt{A} dB_{k,t}
\ee
where $B_{k,t}$ is again a standard Brownian motion. 
Notice that $V$ can model a self-potential as well as an interaction potential.
To obtain the new process applying a transformation ${\cal R}$ of the
type 
$({\cal R}(\omega))_t= U \omega_t$, we remind that if $X$ is multivariate normal distributed with mean zero
and covariance matrix $A$, then
for any $d\times d$ matrix $U$, $UX$
is multivariate normal distributed with mean zero
and covariance $UAU^t$.
As a consequence if $Y_{k,t}= UX_{k,t}$ then
\be
dY_{k,t}= U F(U^{-1} Y_{k,t}) dt + U \nabla_k V(U^{-1} Y_t) dt + \sqrt{UAU^t} dB_{k,t}
\nonumber
\ee
\col{where} we denote $U^{-1}Y_t$ the collection $U^{-1}Y_{k,t}$ $\forall k$.
For \col{the process $Y_t$} to be absolutely continuous w.r.t. the \col{$X_{t}$ process},
we need that
$
UAU^t=A.
$
Moreover, 
assuming that the potential is invariant under the transformation $U$, i.e. $V(Ux)= V(x)$,  
the process $Y_{k,t}$ satisfies
\be\label{transpr}
dY_{k,t}= U F(U^{-1} Y_{k,t}) dt + \nabla_k V( Y_t) dt + \sqrt{A} dB_{k,t}.
\ee
Thus, the process \col{whose paths' distribution} 
is invariant under the transformation ${\cal R}$ 
is 
\be\label{invpr}
dZ_{k,t}= \nabla_k V(Z_t) dt + \sqrt{A} dB_{k,t}
\ee
\col{By using }the Girsanov formula \cite{SV} 
\col{one can compute $dP_X=e^HdP_Z$, i.e. the relative density between} 
the path space measure $P_X$ of the process 
\col{\eqref{origpr}}
and the path space measure $P_Z$ of the process \eqref{invpr}.
Analogously the measure $P_Y$ 
of the process \eqref{transpr} and the measure 
$P_Z$ are related by 
$d(P_X\circ {\cal R})=dP_Y=e^{H\circ {\cal R}^{-1}}dP_Z$. Hence, 
if we denote $\tilde U$ and $\tilde A$ as the $N\times(d\times d)$ block diagonal 
matrices with elements $U$ and $A$ respectively on the diagonal and 
$\tilde F$ as the $N\times d$ column vector consisting in copying $N$ times the vector $F$, 
then by applying \eqref{alt} we get 
\beq
(H\circ {\cal R}^{-1}&-& H )(\omega)\label{HRHgen}
=
\int_0^T \tilde A^{-1}(\tilde U\tilde F(\tilde U^{-1}\omega_t)- \tilde F(\omega_t))   d\omega_t\nonumber
\\
&-&\int_0^T \tilde A^{-1}(\tilde U\tilde F(\tilde U^{-1}\omega_t)- \tilde F(\omega_t)) \cdot \nabla V(\omega_t)dt\nonumber
\\
&-&
\frac12\int_0^T (\tilde U\tilde F(\tilde U^{-1}\omega_t) \cdot \tilde A^{-1}(\tilde U\tilde F(\tilde U^{-1}\omega_t) dt \nonumber
\\
&+& 
\frac12\int_0^T \tilde F(\omega_t) \cdot  \tilde A^{-1} \tilde F(\omega_t) dt.
\eeq
In addition, if the force $F$ is constant we find (we put $\omega_0=0$)
\beq\label{HRH}
(H\circ {\cal R}^{-1}-H) (\omega)
=  A^{-1}( U F- F) \cdot Q_T(\omega)
\eeq
with
$
Q_T(\omega)= \sum_{k=1}^N \Big( \omega_{k,T}- \int_0^T\nabla_k V(\omega_t) dt \Big).
$
\col{Furthermore, by choosing $\phi (\omega)=\lam \cdot Q_T(\omega)$,
with $\lambda\in\R^d$},
we have that
$(\phi\circ {\cal R})(\omega)= \lambda \cdot UQ_T(\omega)= U^t\lambda \cdot Q_T(\omega)$ and 
from \eqref{fluctgen} and \eqref{HRH} we get
$
\E(e^{U^t\lambda \cdot Q_T})=\E( e^{(\lambda + A^{-1}(UF-F) \cdot  Q_T}).
$
By defining $Z_T(\lambda):=\E(e^{\lambda \cdot Q_T})$ it readily follows that 
\be\label{isoho}
Z_T (\lambda)= Z_T\Big((U^{-1})^t(\lambda + A^{-1}F)-A^{-1}F\Big)
\ee
which is the analogous relation to the previously found for the discrete setting \eqref{afr-finite-vol}. Notice 
that if $A$ is the diagonal matrix with elements $A_{s,s}=2a_s$, $F=A E$, $V=0$ and \col{$\mu^{BM}(\lam):=\lim_{T\to \infty} (TL^d)^{-1}\ln Z_T(\lam)$} 
we recover from the above equation the previous result \eqref{afr-infinite-vol} for the anisotropic Brownian motion.

\medskip

\section{Conclusions}
In this work we have derived a {\em spatial fluctuation theorem} (SFT) for interacting particle systems 
and interacting diffusions driven out of equilibrium by an external field. It has been proved in 
both cases that the SFT can be traced back to an invariance property of the microscopic 
path space \col{measure} 
under spatial transformations. Remarkably, this result holds for finite time in the case of interacting diffusions \eqref{isoho}, with $UAU^t=A$ and $V(Ux)=V(x)$, as well as for finite volume 
for interacting particle systems \eqref{afr-finite-vol}. In the latter, the spatial transformations 
yielding the SFT correspond to discrete symmetries associated with the underlying lattice geometry of the system at hand. 
As the system linear size increases these transformations become continuous.

The SFT gives a new perspective on how the microscopic symmetries of the system are reflected at the fluctuating level. Whereas the 
standard fluctuation theorem is based on time reversal-symmetry of the microscopic dynamics, the SFT proves that the spatial symmetries have also a word to say. 
\col{It is} worth emphasizing that 
from the SFT new hierarchies for the currents cumulants and for the non-linear response can be obtained
\col{\cite{IFR,AFR}}. For more details see section 6.2 of the J. Stat. Phys. paper of \cite{IFR}, where these hierarchies are derived 
from the SFT,  $\mu(\lam) = \mu\Big((U^{-1})^t (\lam + E) - E\Big) $,
when the transformation $U$ is a rotation. To obtain the hierarchical cumulants relations that has to consider the limit of infinitesimal rotations, 
namely $U={\mathbb I} + \Delta\theta {\cal L}$, with ${\mathbb I}$ the identity matrix and ${\cal L}$ any generator of 
$d$-dimensional rotations. For instance,
{to the lowest order these hierarchies imply Onsager's reciprocity symmetries and Green-Kubo relations for the linear response coefficients, with the additional prediction that,  the linear response matrix is in fact proportional to the identity.} 
In addition, the bare extension to more than one dimension of the standard fluctuation theorem  only allows to compare one spatial direction and its reverse. It is within the SFT that one can relate probabilities of observing two current values in arbitrary different spatial directions
(see \cite{expIFR} for the experimental measurements of   
the current fluctuations of a self-propelled rod immersed in a sea of spherical beads). 
From the experimental point of view, as  the standard fluctuation relation leads to the 
the study of free-energy differences in terms of work distributions (see e.g. \cite{hummer2001,andrieux2007}), 
one can expect that SFT contains further information 
on the distribution of the mechanical work.
The study of the full Langevin dynamics (including inertia) still remains as an open problem, as well as systems subject to noise that is non-Gaussian.

\section*{Acknowledgements}
The authors acknowledge financial support from the Italian
Research Funding Agency (MIUR) through FIRB project, grant n. RBFR10N90W.


\end{document}